# Epitaxial Growth of Large-area Bilayer Graphene on Ru(0001)

Yande Que,[1] Wende Xiao,[1,a)] Xiangmin Fei,[1] Hui Chen,[1] Li Huang,[1,2] S.X. Du[1] and H.-J. Gao[1,a)]

[1]Institute of Physics, Chinese Academy of Sciences, Beijing 100190, China

[2]Institute of Chemistry, Chinese Academy of Sciences, Beijing 100190, China

**Abstract**

Large-area bilayer graphene (BG) is grown epitaxially on Ru(0001) surface and characterized by low temperature scanning tunneling microscopy. The lattice of the bottom layer of BG is stretched by 1.2%, while strain is absent from the top layer. The lattice mismatch between the two layers leads to the formation of a moiré pattern with a periodicity of ~21.5 nm and a mixture of AA- and AB-stacking. The $\sqrt{3} \times \sqrt{3}$ superstructure around atomic defects is attributed to the inter-valley scattering of the delocalized π-electrons, demonstrating that the as-grown BG behaves like intrinsic free-standing graphene.

a) To whom correspondence should be addressed. Tel.: +86-10-82648035, Fax: +86-10-62556598, Electronic mail: wdxiao@iphy.ac.cn and hjgao@iphy.ac.cn

Epitaxial graphene on transition metal substrates has attracted intense interest in the past few years, driven by the unique opportunities to fabricate large-area uniform graphene layers with low defect density, which is crucial for many applications in future devices. For instance, millimeter-size, high-quality, single-crystalline graphene has been epitaxially grown on Ru(0001).[1,2] Epitaxial graphene on Ru(0001) could exhibit unique structural and electronic properties, such as quantum dots effect,[3] buffer layer effect and template effects for molecular assembly.[4-9] However, previous investigations mainly focused on monolayer graphene (MG). The growth and physical properties of bilayer graphene (BG) on Ru(0001) have rarely been addressed.[10-13] BG grown on Ru(0001) can exhibit ordered moiré pattern with a periodicity of ~3 nm, due to the lattice mismatch between the bottom layer graphene and Ru(0001) substrate. Although the epitaxial growth of graphene on Ru(0001) can be controlled layer-by-layer,[10,14] it is still a great challenge to grow large-area high-quality BG. Previous work on the structural properties of BG/Ru(0001) focused on small BG islands with lateral sizes of a few tens of nanometers.[10,13] In particular, only a few patches of ordered moiré pattern can be visible in such BG islands, suggesting the presence of abundant defects in the bottom layer of BG. Moreover, the structure of BG/Ru remains unclear. It is not clear whether the lattice constants of the two layers of BG are identical, how the two layers of BG are stacked, with or without a twisted angle, in AA or AB fashion. Recently, these issues were investigated via theoretical calculations based on density functional theory (DFT).[15,16]

In this letter, we report on large-area BG grown on Ru(0001), as characterized by low temperature scanning tunneling microscopy (LT-STM). Two types of moiré superstructures are found in the BG/Ru(0001) – one with a periodicity of ~3 nm arising from the lattice mismatch between the bottom layer graphene and Ru(0001), the other with a periodicity of ~21.5 nm due to the parallel alignment of two graphene layers with a slight lattice mismatch.

Our experiments were carried out in an ultrahigh vacuum LT-STM system (Unisoku) with a base pressure of $1 \times 10^{-10}$ mbar, equipped with an ion sputtering gun and electron-beam heater for surface cleaning. The Ru(0001) substrate was cleaned by repeated cycles of ion sputtering using Ar$^+$ with energy of 1 keV, annealing at 1400 K and oxygen exposure at 1200 K ($5 \times 10^{-7}$ mbar, 5 min). Prior to the growth of graphene, the surface cleaning of the Ru(0001) substrate was checked by STM measurements. It is noteworthy that the surface cleaning is essential for growth of large-area high-quality BG, as a contaminated surface with impurities often leads to the formation of BG with small patches of ordered moiré pattern. Large-area BG was grown by exposing the clean Ru(0001) substrate to 100 L ethylene ($1 \times 10^{-6}$ mbar, 100 s) at 1400 K (hereinafter referred to as the *growth temperature*), followed with cooling down to room temperature with a rate of ~60 K/min. Unlike the growth temperature of ~1100 K for MG in our previous work,[2] here the elevated growth temperature of 1400 K increases carbon solubility in bulk Ru, favoring the growth of large-area high-quality BG. STM images were acquired in constant-current mode with electrochemically etched tungsten tips at ~4.2 K, and all given voltages refer to the sample.

Figure 1(a) shows a typical STM image of the as-prepared BG epitaxially grown on Ru(0001). An ordered array of bright spots can be clearly seen. These bright spots form a triangular lattice with a lattice length of ~3 nm, similar to that of the moiré pattern of MG grown on Ru(0001).[2,14] Thus, the formation of this moiré pattern is due to the lattice mismatch between Ru(0001) and the bottom layer of BG, indicating that the top layer covers the bottom layer like a carpet. Similar to the case of MG/Ru(0001), each unit cell of this moiré pattern includes three different regions – atop, fcc and hcp. The carbon atoms of the atop regions of the bottom layer graphene are seated on top of the ruthenium atoms of the Ru(0001) substrate, while those of the fcc and hcp regions are stacked at the hollow fcc and hcp sites of the substrate, respectively.[17] Interestingly, the apparent heights of these bright spots (atop regions) exhibit a periodic modulation, leading to the formation of an additional moiré pattern with a much larger periodicity. The white rhombus in Fig. 1(a) indicates the unit cell. Line profile analysis (Fig. 1(b)) clearly shows the formation of two moiré patterns with periodicities of 2.97 ±0.03 nm and 21.46 ±0.21 nm. The co-existence of two moiré patterns with different periodicities can also be distinguished from fast Fourier transformation (FFT) analysis. As seen in the inset of Fig. 1(a), the outside set of spots corresponds to the reciprocal lattice of the moiré pattern with a periodicity of ~3 nm, while the inner set of spots corresponds to the reciprocal lattice of the moiré pattern with a periodicity of ~21.5 nm.

As the longest periodicity of possible moiré patterns is ~ 3 nm for MG on Ru(0001),[18] the additional moiré pattern with a periodicity of ~21.5 nm can only be originated from the lattice mismatch of the two layers of BG. To unveil the physical origin of this moiré pattern, we acquired atomic resolution STM images of BG grown on Ru(0001). Figure 2(a) shows a typical atomically resolved STM topography of BG. The lattice constant of the top layer of the graphene was measured to be 2.46 ±0.02 Å, akin to that of free-standing MG.[19] This behavior suggests that no strain is built up in the top layer, due to the weak interlayer coupling. The honeycomb lattice of the top layer graphene can be resolved (Fig. 2(a) and (b)). We note that for MG grown on Ru(0001) the symmetry between the A- and B-sublattices (AB-symmetry) in the atop regions is essentially preserved, as the carbon atoms in these regions are located on top of the ruthenium atoms. Meanwhile, the AB-symmetry of MG is broken in the fcc and hcp regions, as the carbon atoms of the A- and B-sublattices are stacked at different sites with respect to the substrate. Thus honeycomb lattices can usually be resolved by STM measurements in the atop regions, corresponding to two sublattices, whereas only triangular lattices can be resolved in the fcc and hcp regions,[11] corresponding to one of the sublattices. Therefore, the observation of honeycomb lattice on BG/Ru(0001) again evidences that the top layer graphene is nearly free-standing. This is rather different from the bottom layer of BG, which exhibits n-doped semiconductor due to the strong coupling between bottom layer graphene and the Ru(0001) substrate.[20,21] It is known that MG grown on Ru(0001) can be described by a model with 12 ×12 unit cells of graphene sitting on 11 ×11 unit cells of Ru (12 ×11 model),[17] or by a more precise one with 25 ×25 unit cells of graphene sitting on 23 × 23 unit cells of Ru (25 × 23 model).[20] According to these models, the <10$\bar{1}$0> directions of the bottom layer graphene and Ru(0001) substrate are parallel. The lattice length of the bottom layer graphene is stretched by ~1%.[17] As no strain is present in the top layer graphene, we therefore propose that the formation of the moiré pattern with a periodicity of ~21.5 nm is due to the lattice mismatch between the bottom and top layers of BG.

The lattice constant of the bottom layer graphene can be precisely determined by the relationship between reciprocal vectors $\mathbf{k}_{bottom} = \mathbf{k}_{top} - \mathbf{k}_{moiré}$, where $\mathbf{k}_{bottom}$, $\mathbf{k}_{top}$ and $\mathbf{k}_{moiré}$ denote the reciprocal lattice vectors of the bottom layer graphene, the top layer graphene and the moiré pattern with a periodicity of 21.5 nm, respectively. From this equation, the lattice constant of the bottom layer graphene was derived to be 2.49 ±0.02 Å, stretched by 1.2% with respect to the free-standing graphene lattice of 2.46 Å. Recently, Martoccia *et al.* reported a stretched strain of ~1 % in MG on Ru(0001) based on surface x-ray diffraction measurements and DFT calculations using the 25 × 23 model.[20] Using the 12 × 11 model, a stretched strain of 0.7 % ~ 0.8% in MG on Ru(0001) was revealed via DFT calculations.[2,17] Our result is in good agreement with previous reports. It is noteworthy that the moiré pattern with a periodicity of ~21.5 nm cannot be reproduced via lattice twisting of the bottom and top layer graphene, according to the same equation.

Furthermore, the lattice constant of the Ru(0001) surface can be precisely calculated by the similar relationship between reciprocal vectors $\mathbf{k}_{Ru(0001)} = \mathbf{k}_{bottom} - \mathbf{k}_{moiré}$, where $\mathbf{k}_{Ru(0001)}$, $\mathbf{k}_{bottom}$ and $k_{moiré}$ denote the reciprocal lattice vectors of the Ru(0001) surface, the bottom layer graphene, and the moiré pattern with period of 2.97 ± 0.03 nm, respectively. From this equation, a lattice constant of 2.72 ± 0.02 Å is derived for the Ru(0001) surface. This value is in line with that of bulk Ru(0001),[22] suggesting that the Ru(0001) surface might be free of strain.

Figure 2(d) shows a structural model of a section of the unit cell of the long-periodicity moiré pattern of BG grown on Ru(0001). For simplicity, the Ru(0001) substrate is not shown. The moiré pattern with a periodicity of ~3 nm due to the lattice mismatch between the bottom layer graphene and Ru(0001) is also omitted. The $<10\bar{1}0>$ directions of both graphene layers are parallel, whereas the lattice of the bottom layer is stretched by 1.2% with respect to that of the top layer. It is clearly seen that the stacking of the two layers is continuously varied from the AA to AB (Bernal type) fashion in each unit cell. This is indeed observed in atomic-resolution STM images shown in Fig. 2. Figure 2(b) illustrates a honeycomb lattice, suggesting that the A- and B-sublattices of the top layer graphene of this region are symmetric. Thus the two layers are AA-stacked in this region. Meanwhile, the triangular lattice shown in Fig. 2(c) indicates that AB-symmetry of graphene lattice is broken and the two layers are AB-stacked in this region, similar to that of graphite.[23] Recently, Peng and Ahuja[15] studied the structural and electronic properties of BG grown on Ru(0001) by theoretical calculations, finding that the bottom layer graphene is stretched due to the graphene-substrate interaction, whereas the top layer graphene is nearly free-standing. The lattice mismatch between the two layers leads to the formation of an additional moiré superstructure with large periodicity and a gradual change from AB- to AA- stacking in each unit cell of this moiré superstructure, in line with our results.

It is well known that surface state electrons scattered from atomic defects give rise to quantum interference patterns in the electron density, which can be directly imaged by STM. From the quantum interference patterns, one can deduce the electronic structures.[24-27] For free-standing graphene, the inter-valley scattering of the delocalized π-electrons around an atomic defect usually leads to the formation of √3 × √3 superstructure with respect to the graphene lattice. Such quantum interference at atomic scale is a fingerprint of electron π states close to the Fermi level.[26] As MG is strongly coupled with the substrate and exhibits an n-doped semiconductor feature, such √3 × √3 superstructures cannot be observed in MG on Ru(0001). For BG on Ru(0001), the top layer graphene is nearly free-standing and such superstructures can be clearly seen. Figure 3(a) presents a typical STM image after argon ion sputtering. The protrusions that appear much brighter than the graphene lattices are due to the electron scattering around the atomic defects.[28] FFT analysis (Fig. 3(b)) reveals that the superstructure around defects can be described as √3 × √3 superstructure with respect to the graphene lattice, akin to the quantum interference patterns found in MG on Pt(111),[29] MG on SiC(0001)[26] and Si-intercalated MG on Ru(0001), where the MG exhibit nearly free-standing feature.[30] The appearance of the √3 × √3 superstructure around atomic defects demonstrates that the as-grown BG behaves like intrinsic free-standing graphene.

In summary, we have grown large-area BG on Ru(0001) and observed, by LT-STM, a moiré pattern with a periodicity of ~21.5 nm with a mixture of AA- and AB-stacking. This moiré pattern originates from the BG's internal lattice mismatch – the bottom layer is stretched by 1.2% due to its strong interaction with the substrate, while the top layer is nearly free-standing. The appearance of the √3 × √3 superstructure around atomic defects arises from the inter-valley scattering of the delocalized π-electrons, showing that the as-grown BG behaves like intrinsic free-standing graphene. Our work might shed light on the growth of large-area high-quality bilayer graphene on transition metals for potential applications.

**Acknowledgments**

The authors gratefully acknowledge financial support from MOST (Nos. 2013CBA01600 and 2011CB932700), NSFC, and CAS.

**Figure Captions**

FIG. 1. (Color online) Superstructures of BG epitaxially grown on Ru(0001). (a) STM image showing a moiré pattern with a periodicity of ~3 nm and an additional moiré pattern with a periodicity of 21.5 nm (sample bias: $U$ = -200 mV; tunneling current: $I$ = 10 pA). The white rhombus indicates the unit cell of this additional moiré pattern. Inset shows the FFT pattern. (b) Line profile along the black line shown in (a). The red dashed line illustrates the modulations of the apparent heights of the atop regions.

FIG. 2. (Color online) (a) High-resolution STM image of BG grown on Ru(0001) ($U$ = -30 mV, $I$ = 10 pA). (b, c) Close-ups showing the honeycomb and triangular lattices of the regions marked with blue solid and black dashed squares in (a), respectively. (d) Structural model of the long-periodicity moiré pattern of BG grown on Ru(0001). For simplicity, the Ru(0001) substrate and the moiré pattern with a periodicity of ~3 nm are not shown. The $<10\bar{1}0>$ directions of both graphene layers are parallel, whereas the lattice of the bottom layer is stretched by 1.2% with respect to that of the top layer. The two layers of BG are AA- and AB-stacked in the regions marked with blue solid and black dash circles, respectively.

FIG. 3. (Color online) (a) Atomic-resolution STM image showing the √3 × √3 superstructure around the atomic defects of BG grown on Ru(0001) after ion irradiation ($U$ = -200 mV, $I$ = 10 pA). (b) FFT pattern of (a). The outer (solid circles) and inner (dashed circles) sets of spots are assigned to the graphene lattice and the √3 × √3 superstructure, respectively.

**Figures**

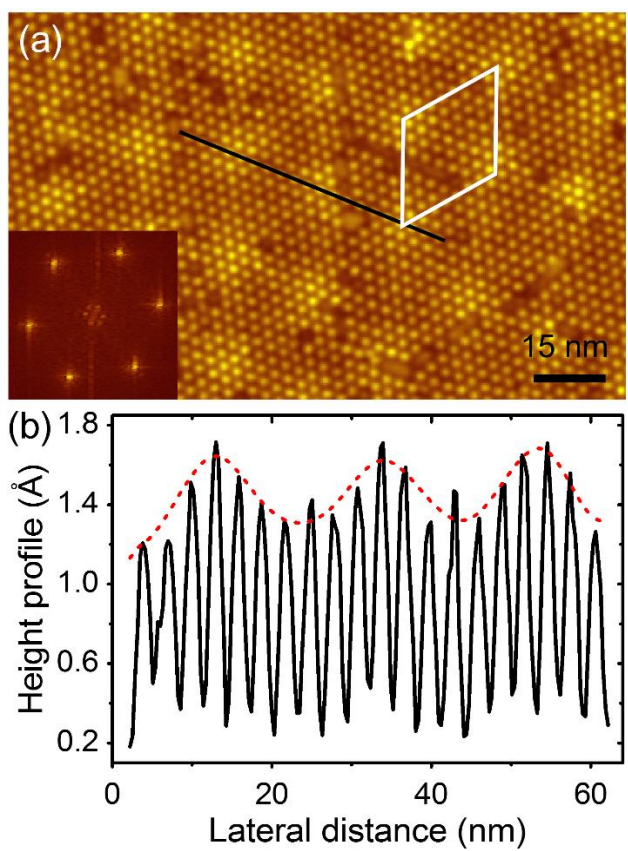

**Fig. 1**

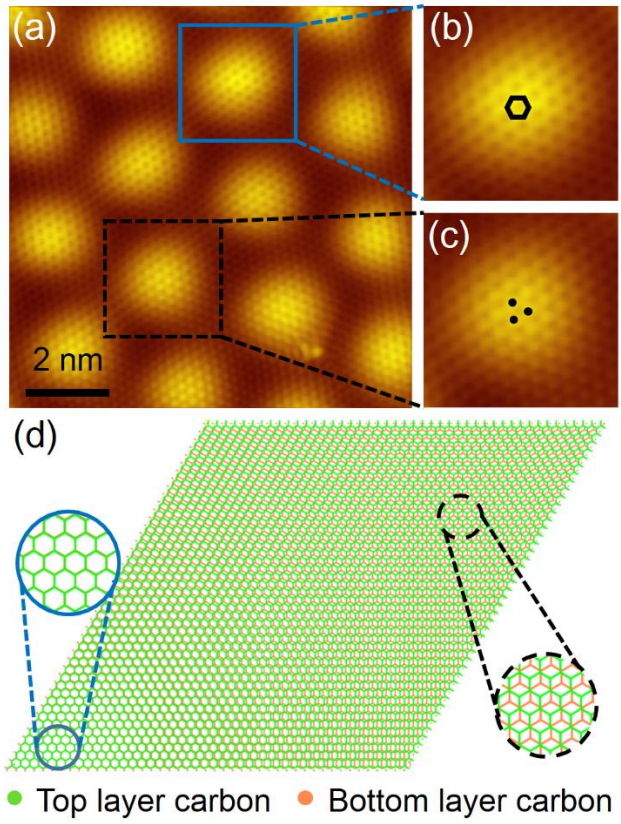

**Fig. 2**

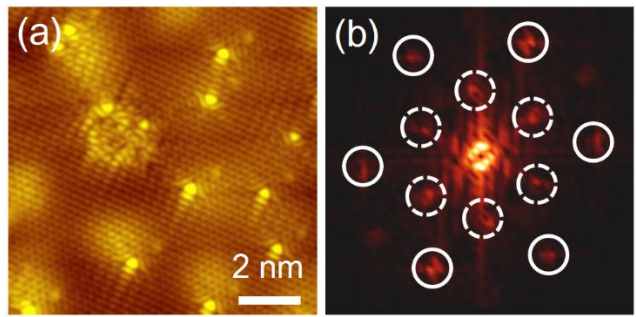

**Fig. 3**